\documentstyle [12pt] {article}
\begin{document}

\def\be{\begin{eqnarray}}
\def\en{\end{eqnarray}}
\def\L{{\cal L}}
\def\tr{{\rm Tr}}
\def\Q{{\cal Q}}
\def\non{\nonumber}
\def\kl{K_L\to \pi^+\pi^-\gamma}
\def\ks{K_S\to\pi^+\pi^-\gamma}
\def\kp{K^+\to\pi^+\pi^0\gamma}
\def\km{K^-\to\pi^-\pi^0\gamma}
\def\kpm{K^\pm\to\pi^\pm\pi^0\gamma}
\def\ep{\varepsilon}
\def\im{ {\rm Im}}
\def\re{{\rm Re}}
\def\ra{\rangle}
\def\la{\langle}
\def\pr{{\sl Phys. Rev.}~}
\def\prl{{\sl Phys. Rev. Lett.}~}
\def\pl{{\sl Phys. Lett.}~}
\def\np{{\sl Nucl. Phys.}~}
\def\zp{{\sl Z. Phys.}~}

\font\el=cmbx10 scaled \magstep2
{\obeylines
\hfill ITP-SB-93-36
\hfill IP-ASTP-22-93
\hfill July, 1993}

\vskip 2.0 cm

\centerline {{\el Radiative Kaon Decays $K^\pm\to\pi^\pm\pi^0\gamma$ and Direct
CP Violation}}

\medskip
\bigskip
\medskip

\centerline{\bf Hai-Yang Cheng}

\medskip
\centerline{ Institute of Physics, Academia Sinica}
\centerline{Taipei, Taiwan 11529, Republic of China}

\centerline{and}

\medskip
\centerline{ Institute for Theoretical Physics, State University of
New York}
\centerline{Stony Brook, New York 11794, USA}

\bigskip
\bigskip

\centerline{\bf Abstract}
\bigskip
   It is stressed that a measurement of the electric dipole amplitude for
direct photon emission in $\kpm$ decays through its interference with inner
bremsstrahlung is important for differentiating among various models. Effects
of amplitude CP violation in the radiative decays of the charged kaon
are analyzed in the Standard Model in conjunction with
the large $N_c$ approach. We point out that gluon and electromagnetic
penguin contributions to the CP-violating asymmetry between the Dalitz plots
of $\kpm$ are of equal weight. The magnitude of CP asymmetry ranges from
$2\times 10^{-6}$ to $1\times 10^{-5}$ when the photon energy in the kaon rest
frame varies from 50 MeV to 170 MeV.

\pagebreak

  In a recent paper [1] we have studied CP violation in the radiative kaon
decay $\kl$. We conclude that the direct CP-violating effect originating from
the electromagnetic penguin diagram is only of order ($10^{-3}-10^{-4})
\epsilon$, depending on the region of the Dalitz plot under consideration. On
the contrary, it has been advocated
that direct CP-violating asymmetry in the radiative decays $K^\pm\to\pi^\pm
\pi^0\gamma$ defined by
\be
\Delta_\Gamma=\,{\Gamma(\kp)-\Gamma(\km)\over \Gamma(\kp)+\Gamma(\km)},
\en
arising from the same electromagnetic penguin mechanism can be large enough for
experimental interest; explicitly, $\Delta_\Gamma\leq 9\times 10^{-4}$
is obtained in Ref.[2]. If this estimate is correct, it will be on the verge
of the capability of the $\phi$ factory DA$\Phi$NE [3], and could be detected
at future high-statistics facilities. The purpose of this short Letter is to
re-examine this CP-violating effect in the Standard Model in conjunction
with the $1/N_c$ approach.

   The general amplitude of the decay $K^+(k)\to\pi^+(p_+)\pi^0(p)\gamma(q,\ep)
$ is of the form
\be
A(\kp) &=& -eA(K^+\to\pi^+\pi^0)\left({p_+\cdot \ep\over p_+\cdot q}-{p
\cdot \ep\over p\cdot q}\right)e^{i{\delta}^2_0} \\
&& +M[ie\epsilon_{\mu\nu\rho\sigma}p^\mu_+ p^\nu q^\rho\ep^\sigma]e^{i{\delta}
^1_1}+Ee[(p_+\cdot\ep)(p\cdot q)-(p\cdot\ep)(p_+\cdot q)]e^{i{\delta}^1_1},
\non
\en
where we have included the isospin phase shifts $\delta^I_J$ with $I$ being
the isospin of the two pions and $J$ the total angular momentum, which are
necessary for generating the decay-rate CP asymmetry. To the leading
multipole expansion, the direct emission (DE) amplitude $M$ corresponds to
a magnetic dipole (M1) transition, while $E$ an electric dipole (E1)
transition. From time to time this decay mode has received a constant
attention both theoretically [2-14] and experimentally [15-17]. Outside of the
framework of chiral perturbation theory (ChPT), various techniques such as
the short-distance effective weak Hamiltonian, the vector-meson-dominance
model and current algebra have been
employed to study the DE of $\kp$ decay. These methods are plagued with some
fundamental problems. For example, the short-distance effective Hamiltonian
utilized in Refs.[5,7] does not explicitly couple to the external photon field.
Hence, after the usage of the factorization approximation,
one has to appeal to the soft-pion theorem to evaluate the matrix
element $\la\pi^0\gamma|\bar{s}\gamma_\mu(1-\gamma_5)u|K^+\ra$, for instance.
However, it has been shown [8] that the soft-pion technique is no longer
applicable to the magnetic transition amplitude as in the case of $\pi^0\to
\gamma\gamma$. Also, it is known that the short-distance effective weak
Hamiltonian is far from being adequate to describe the kaon $\Delta I={1\over
2}$ rule.

 In ChPT, the most general $p^4$ CP-invariant $\Delta S=1$ non-anomalous
electroweak chiral Lagrangian with one external photon field which satisfies
the constraints of chiral and $CPS$ symmetry has the expression [18]
\begin{eqnarray}
\L^{\Delta S=1}_{\rm non-anom} &=& i\left( {2\over f_\pi^2}\right)g_8eF^{\mu
\nu}[\,\omega_1\tr(\lambda_6 L_\mu L_\nu Q)+\omega_2\tr(\lambda_6 L_\nu Q L_
\mu)  \\   &+& \omega_3\tr(\lambda_6 UR_\mu R_\nu QU^\dagger)+\omega_4
\tr(\lambda_6UQR_\mu R_\nu U^\dagger)+\omega_5\tr(\lambda_6UR_\nu QR_\mu
U^\dagger)\,] \non
\end{eqnarray}
for normal intrinsic parity transitions, while anomalous Lagrangian terms
for the odd intrinsic parity sector are [10,12]
\begin{eqnarray}
\L^{\Delta S=1}_{\rm anom} &=& ia\left( {2\over f_\pi^2}\right)g_8e\tilde{F}^{
\mu\nu}\tr(QL_\mu)\tr(\lambda_6 L_\nu)  \non \\
&+& ib\left( {2\over f_\pi^2}\right)g_8e\tilde{F}^{
\mu\nu}\tr(QR_\mu)\tr(\lambda_6 L_\nu)  \non \\
&+& ic\left( {2\over f_\pi^2}\right)g_8e\tilde{F}^{
\mu\nu}\tr\left(\lambda_6[UQU^\dagger,~L_\mu L_\nu]\right),
\end{eqnarray}
where $\tilde{F}_{\mu\nu}\equiv \epsilon_{\mu\nu\alpha\beta}F^{\alpha\beta}$,
$Q={\rm diag}({2\over 3},-{1\over 3},-{1\over 3})$, $L_\mu\equiv (D_\mu
U)U^\dagger$ with $D_\mu U=\partial_\mu U-ieA_\mu[Q,~U]$, $R_\mu\equiv U^
\dagger(D_\mu U)$, and
$U = \exp\left(2i{\phi\over f_\pi}\right)$ with $f_\pi=132\,{\rm MeV},~
\phi\equiv{1\over\sqrt{2}}\phi^a\lambda^a$.
In Eqs.(3) and (4), $g_8$ is the octet weak coupling constant appearing in
the lowest order CP-invariant $\Delta S=1$ weak chiral Lagrangian
$\L^{(2)}_W=-g_8\tr(\lambda_6L_\mu L^\mu)$ and is fixed to be [19]
\begin{eqnarray}
g_8=-0.26\times 10^{-8}m^2_K
\end{eqnarray}
from the experimental measurement of $K^0\to\pi\pi$ decay rates. The coupling
constants $\omega_i,~a,~b,~c$ depend on the choice of the
renormalization scale $\mu$ as divergences of chiral loops are absorbed by the
counterterms which have the same structure as $\L^{\Delta S=1}_{\rm non-anom}$
and $\L^{\Delta S=1}_{\rm anom}$. Therefore, those coupling constants in
principle can be determined only empirically from various low-energy hadronic
processes. However, in the limit of large $N_c$ ($N_c$ being the number of
quark color degrees of freedom), these couplings become $\mu$ independent and
are theoretically manageable at least to the zeroth order of $\alpha_s$ [10].
It is found that in the large-$N_c$ approach [10]
\begin{eqnarray}
\omega_1=\omega_2={N_c\over 12\pi^2},~~~\omega_3=\omega_4=\omega_5=0,~~~
a=2b=4c={N_c\over 12\pi^2}.
\end{eqnarray}
Note that the couplings $a,~b$ and $c$ are determined
by chiral anomalies and hence are free of gluonic corrections in the large
$N_c$ limit.

   There are two different contributions to the direct emission of $\kpm$:
contact-term contributions induced by $\L^{\Delta S=1}_{\rm non-anom}$
and $\L^{\Delta S=1}_{\rm anom}$, and three long-distance pole diagrams with
the $\pi\pi\pi\gamma$ vertex governed by the anomalous Wess-Zumino-Witten term.
As shown in Ref.[10], the results are (Since we are working in the leading
order in $1/N_c$ expansion, chiral loops can be neglected.)
\be
E(\kpm) &=& E_{\rm contact}=\,{\sqrt{2}g_8\over \pi^2f_\pi^5}(2), \non \\
M(\kpm) &=& M_{\rm contact}+M_{\rm pole}=\,\mp{\sqrt{2}g_8\over \pi^2f_\pi^5}
(2+3).
\en
The {\it constructive} interference between pole and direct-transition M1
amplitudes for $\kpm$ decays is a prominent feature different from the decay
$\kl$ where a large and {\it destructive} interference for M1 transitions
is required to explain the data [1,10]. Experimentally, the DE rates are
extracted in the charged-pion kinetic energy range of 55 to 90 MeV [15-17].
With this experimental condition, the branching ratio of direct emission is
given by [10]
\be
Br(\kpm)_{\rm DE} &=& 1.32\times 10^5(|E|^2+|M|^2)\,{\rm GeV}^6  \non \\
&=& 2.02\times 10^{-5},
\en
which is in agreement with the experimental values
\be
Br(\kpm)_{\rm DE}=\cases{(1.56\pm 0.35\pm 0.5)\times 10^{-5} & [1972]~
(Ref.[15]),  \cr (2.3\pm 3.2)\times 10^{-5}&[1976]~ (Ref.[16]),   \cr
(2.05\pm 0.46^{+0.39}_{-0.23})\times 10^{-5} & [1987]~ (Ref.[17]).  \cr}
\en
Previous calculations [5,7,8] based on the short-distance effective weak
Hamiltonian predict a smaller branching ratio. This is attributed to the fact
that, as noted in passing, only short-distance corrections to the Wilson
coefficient functions are taken into account in the approach of the effective
weak Hamiltonian, which are not sufficient to explain the $\Delta I={1\over 2}$
rule in kaon decays (see Ref.[19] for a review). We note that although the DE
rate is dominated by magnetic transitions
\footnote{Precisely, $|E/M|^2=0.16$ is predicted in the large-$N_c$ approach.
Most earlier calculations yield even
smaller ratio for $|E/M|^2$. In Ref.[14] this ratio is calculated to be
$5.1\times 10^{-3}$. In ChPT, the coupling constants of $\L^{\Delta S=1}_{\rm
non-anom}$ and $\L^{\Delta S=1}_{\rm anom}$ are expected to be of the same
order of magnitude [see Eq.(6)]. Therefore, unlike the decay $\kl$, it seems
to us that there is no reason to have a severe suppression on the
electric dipole amplitude of DE in $\kpm$ decays.}
due to additional constructive contributions from the pole diagrams, the
E1 contribution is nevertheless not negligible. Experimentally,
the DE electric dipole amplitude can be measured from the interference of
inner bremsstrahlung (IB) with E1 transitions. Thus far, there is only one
experiment (done 2 decades ago) measuring this interference and only
a limit is obtained [15]. Evidently, a measurement of the E1 amplitude is
important for understanding the underlying mechanism of $\kpm$ decays and
their direct CP violation.

   We next turn to examine CP-violating effects in the decays $\kpm$. The
simplest way of observing CP nonconservation is through the measurement of
the decay-rate CP asymmetry parameter $\Delta_\Gamma$ as defined in Eq.(1).
If photon polarization is not measured, there is no interference between E1
and M1 amplitudes. Consequently, when photon polarizations are summed over,
a nonvanishing $\Delta_\Gamma$ must arise from the interference of IB with
the E1 amplitude of DE. Although CP-violating asymmetry has been studied
intensively in late 1960's [20], a modern analysis in the framework of the
Standard Model was carried out only recently in Refs.[2] and [7]. However,
the result is somewhat controversial: While $\Delta_\Gamma$ arising from the
gluon penguin diagram is estimated to be of order $10^{-6}$ by McGuigan and
Sanda [7], it is claimed by Dib and Peccei [2] that a large CP asymmetry
can be induced from the electromagnetic penguin diagram, namely $\Delta_\Gamma
\leq 9\times 10^{-4}$. Naively, it is expected that amplitude CP violation
coming
from the electromagnetic penguin diagram with a photon radiated from the loop
quark or from the $W$ boson is of equal weight as that from
the QCD penguin diagram with a photon emitted from the external quark lines
or from the $W$ boson. Therefore, a resolution of this discrepancy is called
for.

  We will follow Ref.[7] to consider CP asymmetry between the Dalitz plots of
$\kp$ and $\km$ rather than in the total decay rates
\be
\Delta=\,{|A(\kp)|^2-|A(\km)|^2\over |A(\kp)|^2+|A(\km)|^2},
\en
so that a larger CP asymmetry can be obtained in certain particular regions
of the Dalitz plot. Since under CPT invariance
\be
A(\km) &=& -eA(K^-\to\pi^-\pi^0)\left({p_-\cdot \ep\over p_-\cdot q}-{p
\cdot \ep\over p\cdot q}\right)e^{i{\delta}^2_0} \\
&& -M^*[ie\epsilon_{\mu\nu\rho\sigma}p^\mu_- p^\nu q^\rho\ep^\sigma]e^{i{
\delta}^1_1}+E^*e[(p_-\cdot\ep)(p\cdot q)-(p\cdot\ep)(p_-\cdot q)]e^{i{\delta}
^1_1}, \non
\en
it follows that
\be
\Delta=\,{-2|E|z\sin\phi_E\sin(\delta^1_1-\delta^2_0)\over A(K^+\to\pi^+
\pi^0)/m^4_K+2|E|z\cos\phi_E\cos(\delta^1_1-\delta^2_0)+(|E|^2+|M|^2)z^2},
\en
where $E=-|E|e^{i\phi_E}$ [recall that our $E$ is negative; see Eqs.(5) and
(7)] and $z\equiv(q\cdot p_\pm)(q\cdot p)/m^4_K$.

   The main task is to estimate the CP-odd phase $\phi_E$ of the E1 amplitude.
There
are two different contributions to the imaginary part of $E$: one from the
gluon penguin diagram, and the other from the electromagnetic penguin diagram.
 As to the former, following the prescription presented in Ref.[1], we find
in the $1/N_c$ approach that
\be
E=\,{2\sqrt{2}\over \pi^2f_\pi^5}(g_8+ig_8'),
\en
where $g'_8$ is the CP-violating coupling constant appearing in the lowest
order CP-odd $\Delta S=1$ weak chiral Lagrangian $\L_W^-=-ig'_8\tr(\lambda_7
L_\mu L^\mu)$ and is dominated by the short-distance QCD penguin diagram.
Hence,
\be
(\sin\phi_E)_{\rm gluon}=\,{g'_8\over g_8}=\,{\im A_0\over \re A_0},
\en
where $A_0\equiv A(K^0\to\pi\pi(I=0))$, and use of
\be
{A(K_2\to\pi\pi(I=0))\over A(K_1\to\pi\pi(I=0))}=\,i{g'_8\over g_8}
\en
has been made. The calculation of Im$A_0$ in the Standard Model is standard
and is given by [21]
\be
\im A_0=-{G_F\over\sqrt{2}}(\im\lambda_t)y_6\la(\pi\pi)_{I=0}|Q_6|K^0\ra,
\en
where $y_6=\im c_6/\im\tau,~\tau=-\lambda_t/\lambda_u,$ $\lambda_i=V_{is}^*
V_{id}$, $Q_6$ is a penguin operator and $c_6$ is the corresponding Wilson
coefficient. The $K-\pi\pi$ matrix element of $Q_6$ evaluated in the large
$N_c$
approach is known to be [19]
\be
\la(\pi\pi)_{I=0}|Q_6|K^0\ra=-i4\sqrt{3}f_\pi v^2\,{m_K^2-m^2_\pi\over \Lambda
_\chi^2},
\en
where
\be
v=\,{m^2_\pi\over m_u+m_d}=\,{m^2_{K^+}\over m_u+m_s}=\,{m^2_{K^0}\over m_d
+m_s}
\en
characterizes the quark order parameter $\la\bar{q}q\ra$, and $\Lambda_\chi
\sim 1$ GeV is a chiral symmetry breaking scale. Since $\re A_0=i4.69\times
10^{-7}$ GeV [19]
\footnote{The experimental values of $K\to\pi\pi$ amplitudes are usually
expressed in terms of real numbers. However, model calculations show that the
amplitude of $K\to 2\pi$ contains a factor of $i$. This means that Re$A_0$
given in Ref.[19] should be multiplied by a factor of $i$ when compared with
$\la\pi\pi|Q_6|K^0\ra$.}
and
\be
\im(V^*_{ts}V_{td})\simeq s_{13}s_{23}\sin\delta_{13}
\en
in the Chau-Keung parametrization of the quark mixing mixing matrix [22], we
find numerically
\be
(\sin\phi_E)_{\rm gluon}=-8.1\times 10^{-5}\sin\delta_{13},
\en
where uses have been made of $m_s=175$ MeV, $s_{23}=0.044,~s_{13}/s_{23}=0.1$,
and $y_6=-0.057$ for $m_t=150$ GeV [21].

   Following Ref.[1], the DE amplitude of $\kp$ induced by the electromagnetic
penguin diagram is given by
\begin{eqnarray}
A(\kl)_{\rm DE}^{\rm em}=\,iG_F\,{e\over 16\pi^2}\im(V^*_{ts}V_{td})F(x_t)\la
\pi^+\pi^0\gamma|Q_T|K^+\ra,
\end{eqnarray}
with
\begin{eqnarray}
 Q_T &=& i[m_s\bar{s}\sigma_{\mu\nu}(1-\gamma_5)d+m_d\bar{s}
\sigma_{\mu\nu}(1+\gamma_5)d]F^{\mu\nu}, \non \\
 F(x) &=& {(8x^2+5x-7)x\over 12(x-1)^3}-{(3x-2)x^2\over 2(x-1)^4}\ln x,
\end{eqnarray}
and $x_t=m_t^2/M^2_W$. By working out the chiral realization of the tensor
operator $Q_T$ as in Ref.[2] (see also Ref.[1]), we obtain
\begin{eqnarray}
E_{\rm em}=i{G_Fm_s\over 2\sqrt{2}\pi^2f_\pi^2}F(x_t)(s_{13}s_{23}\sin
\delta_{13}).
\end{eqnarray}
It follows from Eqs.(23) and (7) that
\be
(\sin\phi_E)_{\rm em}=\,{-iE_{\rm em}\over E}=\,{1\over 8}\,{G_Fm_sf_\pi^3\over
g_8}F(x_t)(s_{13}s_{23}\sin\delta_{13}).
\en
Numerically,
\be
(\sin\phi_E)_{\rm em}=-6.0\times 10^{-5}\sin\delta_{13},
\en
where we have applied Eq.(5). It is evident that gluon and electromagnetic
penguin contributions to the CP-odd phase of the electric dipole DE amplitude
are equally important, as it should be.

   At this point we would like to comment our work in relation to the study
in Ref.[2]. Dib and Peccei first calculated CP asymmetry for charged kaon decay
into two pions and then applied the CPT relation
\footnote{In principle, one should also include the decay rates of $K\to 3\pi$
decays, namely $\Gamma(K^+\to\pi^+\pi^+\pi^-)+\Gamma(K^+\to\pi^+\pi^0\pi^0)$,
to the l.h.s. of Eq.(26) and their charge conjugate to the r.h.s.}
\be
\Gamma(K^+\to\pi^+\pi^0)+\Gamma(\kp)=\Gamma(K^-\to\pi^-\pi^0)+\Gamma(\km)
\en
to estimate $\Delta_\Gamma$ for $\kpm$ decays. On the contrary, we compute
CP-violating asymmetry directly for the radiative decays of $K^\pm$.
Therefore, the strong-interaction phase difference necessary for generating
CP-odd asymmetry is $\sin(\delta^1_1-\delta^2_0)$ in our case, while it
is $\sin(\delta_\gamma-\delta^2_0)$ in Ref.[2], where $\delta_\gamma$ is the
strong-interaction phase shift for $K^\pm\to\pi^\pm\pi^0$ amplitudes
involving a $\pi\pi\gamma$ intermediate state. Apart from this, it seems to
us that the
numerical discrepancy between the present work and Ref.[2] lies mainly
in the fact that a factor of $1/(4\pi)$ is missing in Eq.(15) of Ref.[2]
for the effective Lagrangian of electromagnetic penguins.
Consequently, $\Delta_\Gamma$ is overestimated by a factor
of $(4\pi)^2$; in other words, the predicted upper bound for $\Delta_\Gamma$
in Ref.[2] should read $5.6\times 10^{-6}$ instead of $9\times 10^{-4}$.
Nevertheless, Dib and Peccei did point
out the importance of the electromagnetic penguin diagram, which is
no longer negligible for $m_t>M_W$. From Eqs.(20) and (25) we see that gluon
and electromagnetic penguin diagrams contribute constructively to $\phi_E$.
As a result,
\be
\sin\phi_E=(\sin\phi_E)_{\rm gluon}+(\sin\phi_E)_{\rm em}=-1.4\times 10^{-4}
\sin\delta_{13}.
\en

    It remains to work out the quantity $z$ defined in Eq.(12). It can be
recast in terms of the variables $x=2k\cdot q/m^2_K$ and $y=2k\cdot
p_+/m^2_K$:
\begin{eqnarray}
z={1\over 4}[x(1-y)-(1-y)^2].
\end{eqnarray}
Note that in the kaon rest frame, $x=2E_\gamma/m_K,~y=2E_+/m_K$. It is easily
seen from Eq.(28) that the maximum $z$ for a given
photon energy $E_\gamma$ in the c.m. is given by
\begin{eqnarray}
(z)_{\rm max}={1\over 4}\left({E_\gamma\over m_K}\right)^2.
\end{eqnarray}
Since $(\delta^1_1-\delta^2_0)\sim 10^\circ$ [15,16] and $A(K^+\to\pi^+\pi^0)
=1.829\times 10^{-8}$ GeV [19], it follows from Eqs.(7), (12), (25) and (29)
that
\be
\Delta(E_\gamma)=\,{0.75\times 10^{-5}\left({E_\gamma\over 100~{\rm MeV}}
\right)^2 \over
1+0.31\left({E_\gamma\over 100~{\rm MeV}}\right)^2 }.
\en
This CP-odd asymmetry ranges from $2\times 10^{-6}$ to $1\times 10^{-5}$
when $E_\gamma$ varies from 50 MeV to its highest value of 170 MeV.

  To conclude, we have shown in the $1/N_c$ approach that the E1 amplitude of
DE in $\kpm$ decays is not negligible. Therefore, a measurement of the
interference of inner bremsstrahlung with electric dipole transitions is
important for differentiating between various models. We also pointed out that
CP-violating asymmetry between the Dalitz plots of $\kpm$ decays receive
equally important contributions from gluon and electromganetic penguin
diagrams. The magnitude of CP asymmetry ranges from
$2\times 10^{-6}$ to $1\times 10^{-5}$ when the c.m. photon energy
varies from 50 MeV to 170 MeV.

  {\it Note added}: After this work was typed, we learned a paper by G. Ecker,
H. Neufeld and A. Pich [CERN-TH-6920/93 and UWThPh-1993-22] in which the
decay $\kp$ is analyzed and a potentially sizeable electric amplitude
interfering with bremsstrahlung is emphasized.
\vskip 4.0 cm
\centerline{\bf Acknowledgments}
\vskip 1.0 cm
The author wishes to thank Prof. C. N. Yang and the
Institute for Theoretical Physics at Stony Brook for their hospitality
during his stay there for sabbatical leave. A correspondence
with A.A. Bel'kov and A. Schaale is acknowledged.
This research was supported in part by the
National Science Council of ROC under Contract No.  NSC82-0208-M001-001Y.

\vskip 1.5 cm

\centerline{\bf REFERENCES}

\bigskip

\begin{enumerate}

\item H.Y. Cheng, ITP-SB-93-32 (1993).

\item C.O. Dib and R.D. Peccei, \pl {\bf B249}, 325 (1990).

\item G. D'Ambrosio, M. Miragliuodo, and P. Santorelli, LNF-92/066 (1992).

\item S.V. Pepper and Y. Ueda, {\sl Nuovo Cimento}, {\bf 33}, 1614 (1964);
S. Oneda, Y.S. Kim, D. Korff, \pr {\bf 136}, B1064 (1964);
M. Moshe and P. Singer, \pl {\bf 51B}, 367 (1974).

\item J.L. Lucio M., \pr {\bf D24}, 2457 (1981).

\item A.A. Bel'kov, Yu.L. Kalinowskii, V.N. Pervushin, and N.A. Sarikov, {\sl
Sov. J. Nucl. Phys.} {\bf 44}, 448 (1986).

\item M. McGuigan and A.I. Sanda, \pr {\bf D36}, 1413 (1987).

\item H.Y. Cheng, S.C. Lee, and H.L. Yu, \zp {\bf C41}, 223 (1988).

\item S. Fajfer, \zp {\bf C45}, 293 (1989).

\item H.Y. Cheng, \pr {\bf D42}, 72 (1990).

\item G. Ecker, H. Neufeld, and A. Pich, \pl {\bf B278}, 337 (1992).

\item J. Bijnens, G. Ecker, and A. Pich, \pl {\bf B286}, 341 (1992).

\item A.N. Ivanov, M. Nagy, and N.I. Troitskaya, {\sl Mod. Phys. Lett.}
{\bf A8}, 1599 (1993).

\item A.A. Bel'kov, A.V. Lanyov, and A. Schaale, DESY 93-060 (1993).

\item R.J. Abrams {\it et al.,} \prl {\bf 29}, 1118 (1972).

\item K.M. Smith {\it et al.,} \np {\bf B109}, 173 (1976).

\item V.N. Bolotov {\it et al.,} {\sl Sov. J. Nucl. Phys.} {\bf 45}, 1023
(1987).

\item G. Ecker, A. Pich, and E. de Rafael, \np {\bf B291}, 692 (1987); \pl
{\bf B189}, 363 (1987); \np {\bf B303}, 665 (1988).

\item H.Y. Cheng, {\sl Int. J. Mod. Phys.} {\bf A4}, 495 (1989).

\item V.G. Soloviev and M.V. Terentev, {\sl Sov. Phys. JETP Lett.}
{\bf 2}, 213 (1965); D. Cline, {\sl Nuovo Cimento} {\bf 36}, 1055 (1965);
\prl {\bf 16}, 367 (1966); G. Costa and P.K. Kabir, \prl {\bf 18}, 429 (1967);
{\sl Nuovo Cimento}, {\bf 41A}, 564 (1967); S. Barshay, \prl {\bf 18}, 515
(1967); N. Christ, \pr {\bf 159}, 1292 (1967).

\item See, e.g. G. Buchalla, A.J. Buras, and M.K. Harlander, \np {\bf B337},
313 (1990).

\item L.L. Chau and W.Y. Keung, \prl {\bf 53}, 1802 (1984).

\end{enumerate}

\end{document}